# BELL INEQUALITIES AND PSEUDO-FUNCTIONAL DENSITIES

Johannes F. Geurdes[1]

**Abstract**

A local hidden variables model with pseudo-functional probability density function restricted to a binary-event probability space is able to reproduce the quantum correlation in an Einstein-Podolsky-Rosen-Bohm-Aharonov experiment.

## 1. Introduction

In physics, Bell inequalities[1] (BI's) are of fundamental importance because they solved[2,3] Einstein's worries about completeness of quantum mechanics.[4] In the mid-fifties, Bohm[5] reformulated Einstein's incompleteness arguments into a correlation between spin-states of spatially separated particles, originally in the singlet state. BI's refer to this situation.

Research, with[6] or without[7-9] BI's, pointed at the inconsistency of adding local hidden variables (LHV's) to quantum mechanics. This raises questions about BI's themselves. If they appear unnecessary they might also not eliminate all LHV's allowed in standard probability theory.[8] Moreover, if this is demonstrated by construction of a valid model, LHV's cannot be inconsistent with quantum correlation.

$$alignl P(\vec{a},\vec{b}) = \ d\lambda \ \rho(\lambda) A(\vec{a},\lambda) B(\vec{b},\lambda).$$

The correlation between measurements of two spins, using local hidden variable(s), $\lambda$, is Here, $\rho(\lambda)$ is the probability density function (PDF). Functions $A(\vec{a},\lambda)$ and $B(\vec{b},\lambda)$, $\vec{a}^2=\vec{b}^2=1$, represent the result of the measurement (ideally $\pm1$) at two distant spin measurement devices.

Because A is independent of parameter vector $\vec{b}$ of the device $D_B$, and vice versa, locality[9] is maintained.

$$/P(\vec{a},\vec{b}) - P(\vec{a},\vec{d})/ + /P(\vec{c},\vec{b}) + P(\vec{c},\vec{d})/ \leq 2.$$

For regular probability densities, the following inequality holds[10]

---

[1]C. van der Lijnstraat 164, 2593 NN Den Haag, The Netherlands

The quantum correlation, $P(\vec{a},\vec{b}) = -(\vec{a} \bullet \vec{b})$, violates the inequality. Hence, a distinction between hidden variable predictions and quantum mechanical predictions is possible.

In this letter we present a classical probabilistic singular LHV's model which meets the

$$\rho = \rho(\lambda) \geq 0, \forall \lambda \in \Lambda,$$

$$< \rho > = 1,$$

$$A = A(\vec{a},\lambda) = +-1, B = B(\vec{b},\lambda) = \pm 1,$$

$$|< \rho A >|= 0, |< \rho B >|= 0,$$

$$< \rho A^2 > = < \rho B^2 > = 1,$$

$$< \rho AB > = -(\vec{a} \bullet \vec{b}),$$

following:

with $<f>= d\lambda f(\lambda)$. An extra condition is that the density is associated with a genuine probability space.

$$P_f \frac{\theta(x)}{x} = \{ \begin{array}{l} 1/x, x > 0 \\ 0, \quad x \leq 0 \end{array}$$

In the probability density, the pseudo-function is used which represents the positive branch of the principal value function.[10] Here, $\theta(x)=1$, when, $x>0$, $\theta(x)=1/2$, when, $x=0$, while, $\theta(x)=0$, when $x<0$.

Moreover, two integrals are crucial. Let us inspect $\lambda \varepsilon [-\beta,+\beta]$, with $\beta=\exp(3^{-3/4}) \approx 1.5507$.

$$\int_{-\beta}^{+\beta} d\lambda \, P_f \frac{\theta(\lambda)}{\lambda} = -_p \int_0^\beta d\lambda overlambda = C = 3^{-3/4},$$

We find that

with $_p$ Hadamard's finite part.[13]  The integral of the product of the pseudo-function and the sign

$$\int_{-\beta}^{+\beta} d\lambda \, P_f \frac{\theta(\lambda)}{\lambda} \, sign(f - \lambda) = {}_{-p}\{ \int_0^f \frac{d\lambda}{\lambda} - \int_f^\beta \frac{d\lambda}{\lambda} \} = -1 \, ,$$

function, defined here as sign(x)=+1, x≥0, while sign(x)= -1, x<0, is
when f=exp((C-1)/2)≈0.7553, C=3$^{-3/4}$≈0.4387.  Note that f<β.  The choice of parameters will become clear later on.

Subsequently, the model is presented and Eq. (3) is verified.  First, the integrations are

$$\prod_{q=1}^3 \sum_{n_q=1}^3 \prod_{\tau=1}^4 \int_{-\beta}^{+\beta} d\,\lambda_\tau \, intfrom_{-\infty}^{+\infty} d\,\chi \prod_{s=1}^2 \int_{-1}^{+1} d\,\eta_s <\bullet\,\bullet\,\bullet> .$$

understood as
Hence, the variables $n_q$, $\lambda_\tau$ and $\eta_s$, in the specified ranges, are postulated.

$$\rho = \{ \quad \frac{1}{4}\prod_{\tau=1}^4 P_f \frac{\theta(\lambda_\tau)}{\lambda_\tau} \frac{1}{\sqrt{2\pi}} \exp(-\chi^2/2) \, , \Gamma = true,$$
$$0, \; \Gamma = false$$

Second, the singular probability density function is specified by
Here, $\Gamma(n_1,n_2,\lambda_1,...,\lambda_4,\chi,\eta_1,\eta_2)$ is true when the following conditions apply.  In the first place, $n_q$ must only run through the *complete* set {1,2,3}, excluding all other sets.  Secondly, $\lambda_\tau$ must only run through the *complete* interval [-β,β], excluding other intervals.  Thirdly, the $\chi$, must run trough the *complete* real interval (-∞,∞) excluding other intervals, subsets of (-∞,∞).  Fourthly, $\eta_s$ must only run through the *complete* set [-1,1], excluding all other intervals.  If one of the associated sets that defines the range of a particular hidden variable is unequal to the one indicated, i.e., to {1,2,3}, [-β,β], (-∞,∞), or, [-1,1], *and/or* the variable lies outside the indicated set, then $\Gamma(n_1,...,\eta_2)$ is false.  Note that it is not unusual that intervals determine a PDF.

In addition, let us specify a delta function for sets X and Y as  δ(X,Y)=1, when X=Y, while δ(X,Y)=0 when X≠Y.  Moreover, the universal event is given by Ξ={1,2,3}$^3$×[-β,β]$^4$×(-∞,∞)×[-1,1]$^2$, with {1,2,3}$^3$={1,2,3}×{1,2,3} ×{1,2,3}, etc, whereby × is the Cartesian product, and { } is the empty set.

From the previous specifications, we see a binary-event probability space associated to ρ.  Accordingly[10] this space is written as (Q,_,P$_\rho$[●]), with _={Ξ,{ }} and P$_\rho$[Q]=<δ(Q,Ξ)ρ>, Q∈_.

Hence additivity, $P(Q)+P(Q^c)=1$, with $Q^c=\Xi\backslash Q$, $(Q\cap Q^c=\{\})$, is warranted in $(Q,\_,P_p[\bullet])$, while the other basic axioms (Ref.10, page 22) also hold for $(Q,\_,P_p[\bullet])$. Hence, an elementary probability space can be associated to the density $\rho$, from which valid probability measures can be obtained. Observe that BI's do not forbid binary probability spaces, such as $(Q,\_,P_p[\bullet])$, to be associated to densities to be used in Eq. (1). Moreover, the discreteness of the spin space coincides with the probability space.

$$A = sign(\,\chi\,)\,sign\{\,\delta_{n_1,n_2}\,a_{n_2}\,sign(f - \lambda_1)\,sign(f - \lambda_2) - \eta_1\},$$

Fourth, in the model the functions A and B are given by

$$B = -sign(\,\chi\,)\,sign\{\,\delta_{n_1,n_3}\,b_{n_3}\,sign(f - \lambda_3)\,sign(f - \lambda_4) - \eta_2\,\},$$

and
with $\delta_{x,y}=1$ when $x=y$, while $\delta_{x,y}=0$ when $x\neq y$, and $A^2=B^2=1$.

$$\frac{1}{\sqrt{2\pi}}\int_{-\infty}^{+\infty} d\chi \exp(-\chi^2/2)\,sign(\,\chi\,) = 0,$$

$$\frac{1}{\sqrt{2\pi}}\int_{\infty}^{+\infty} d\chi \exp(\chi^2/2)\{sign(\,\chi\,)\}^2 = 1.$$

We first may observe that $\rho\geq0$. Second, it follows that $<\rho>=1$. Third, because it follows that $<\rho A>=<\rho B>=0$, while $<\rho A^2>=<\rho B^2>=1$. Fourth, because $\vec{a}^2=\vec{b}^2=1$, the density

$$< \rho AB > = -\sum_{k=1}^{3}\sum_{m=1}^{3}\sum_{n=1}^{3}\delta_{k,m}\,\delta_{k,n}\,a_m\,b_n$$

$$_-\{\int_{-\beta}^{+\beta} d\lambda\; P_f\,\frac{\theta(\lambda)}{\lambda}\,sign(f - \lambda)\,\}^4,$$

entails that we also have

from which it follows that $<\rho AB>=-(a_1b_1+a_2b_2+a_3b_3)=-(\vec{a}\bullet\vec{b})$.

For completeness, the conflict with BI's emerges from operations with absolute signs leading to BI's. In our case we have

$$I = |\int_{-\beta}^{+\beta} d\lambda\, P_f\, \frac{\theta(\lambda)}{\lambda}\, sign(f - \lambda\,)| \leq \int_{-\beta}^{+\beta} d\lambda\, |P_f\, \frac{\theta(\lambda)}{\lambda}| = 3^{-3/4},$$

which is contradictory. Hence, no straightforward path may lead here to BI's.

This concludes the proof that a valid local hidden variables model is possible that remains within the bounds of classical probability theory. Because of this, the model cannot beforehand be dismissed as unphysical.